\documentclass{sobolev}   
\setpg{1}                  

\titl[Testing the variation of fundamental constants]{Testing the variation of fundamental constants 
by astrophysical methods: overview and prospects}

\authors{S.A.\,Levshakov}{S.A.\,Levshakov\aff{1,2}}

\affiliat{
\item  A.F.\,Ioffe Physical-Technical Institute, 194021 St.~Petersburg   
\item St.~Petersburg Electrotechnical University ``LETI'', 197376 St.~Petersburg
}

\email{lev@astro.ioffe.rssi.ru}   

\newcommand{\dmm}{$\Delta\mu/\mu$}
\newcommand{\daa}{$\Delta\alpha/\alpha$}
\newcommand{\kms}{km~s$^{-1}$}
\newcommand{\ms}{m~s$^{-1}$}
\newcommand{\zabs}{$z_{\rm abs}$}

\def\la{\;
\raise0.3ex\hbox{$<$\kern-0.75em\raise-1.1ex\hbox{$\sim$}}\; }
\def\ga{\;
\raise0.3ex\hbox{$>$\kern-0.75em\raise-1.1ex\hbox{$\sim$}}\; }

\begin{document} 

\begin{abstract}
  
By measuring the fundamental constants in astrophysical objects one can test
basic physical principles as space-time invariance of physical laws along with
probing the applicability limits of the standard model of particle physics.
The latest constraints on the fine structure constant $\alpha$
and the electron-to-proton mass ratio $\mu$
obtained from observations at high redshifts and in the Milky Way disk are reviewed. 
In optical range, the most accurate measurements have
already reached the sensitivity limit of available instruments, and further 
improvements will be possible only with next generation of telescopes and receivers.
New methods of the wavelength calibration should be realized to control 
systematic errors at the sub-pixel level.
In radio sector, the main tasks are the search for galactic and extragalactic 
objects suitable for precise molecular spectroscopy as well as high resolution
laboratory measurements of molecular lines to provide accurate frequency standards.
The expected progress in the optical and radio astrophysical observations is
quantified.
  
\end{abstract}

\section{Introduction}\label{sect-1}

The idea that the fundamental physical constants may vary on the cosmological time scale has been discussing 
since 1937, when Milne and Dirac argued about possible variations of the Newton constant $G$ during
the lifetime of the universe~\cite{Mi,Di}.
Currently, the subject of the cosmological variation of fundamental constants is closely related to
emergence considerations of different cosmological models 
inspired by the discovery of late time acceleration of the expansion of the universe~\cite{Ri,Pe}.
The possibility that dimensionless coupling constants such as electron-to-proton mass ratio 
$\mu = m_{\rm e}/m_{\rm p}$
and the fine structure constant $\alpha = e^2/\hbar c$ 
may roll with cosmic time has recently been reviewed in~\cite{Uz,Li,Ma,Mar}. 

The variation of fundamental constants would imply a violation of the Einstein equivalence
principle (EEP), that is, local position invariance (LPI) and local Lorentz invariance (LLI).
In particular, a changing $\alpha$ accompanied by variation in other coupling constants
can be associated with a violation of LLI~\cite{Ko}, and
LPI postulates that the fundamental physical laws are space-time invariant.
The standard model of particle physics (SM) is based on the EEP; thus, 
we can probe the applicability limits of the SM and new types of interactions
by experimental validation of the EEP. 

In spite of some claims that changes in $\alpha$ or $\mu$ were  
marginally detected at high redshifts, 
to date no confirmed variation of dimensionless coupling constants
has been found
on astronomical space-time scales. Below we review current observational
constraints on $\alpha$ and $\mu$ variations
which provide limits on the allowed deviations from the SM and $\Lambda$CDM cosmology.

\section{Basics of the astronomical measurements}\label{sect-2}

Two dimensionless coupling constants $\mu$ and $\alpha$
are of particular interest for astronomical studies since
their fractional changes
\dmm~$= (\mu_{\rm obs} - \mu_{\rm lab})/\mu_{\rm lab}$, and
\daa~$= (\alpha_{\rm obs} - \alpha_{\rm lab})/\alpha_{\rm lab}$
can be accurately measured from spectral line profiles 
of Galactic and extragalactic sources. 

Differential measurements of \dmm\ and \daa\
are based on the comparison of the line centers in the
absorption/emission spectra of cosmic objects and the corresponding laboratory
values. 
It was shown that electro-vibro-rotational lines of H$_2$~\cite{VL} 
and CO~\cite{Sal} have their own sensitivities to $\mu$-variation.
Similarly, each atomic transition is characterized by its individual
sensitivity to $\alpha$-variation~\cite{Dz}.
The dependence of an atomic frequency $\omega$ on $\alpha$ in the
comoving reference frame of a distant object located at redshift $z$
is given by:
$\omega_z = \omega + qx + O(x^2)$, where $x \equiv (\alpha_z/\alpha)^2 - 1$.
Here $\omega$ and $\omega_z$ are the frequencies corresponding to the
present-day value of $\alpha$ and that at a redshift $z$. 
The so-called $q$ factor is an individual parameter for each atomic transition~\cite{Dz}.
If $\alpha_z \neq \alpha$, then $x \neq 0$ 
and the corresponding frequency shift $\Delta\omega = \omega_z - \omega$ is 
${\Delta\omega}/{\omega} = Q{\Delta\alpha}/{\alpha}$,
where $Q = 2q/\omega$ is the dimensionless sensitivity coefficient.

For two lines of the same element with 
the sensitivity coefficients $Q_1$ and $Q_2$, 
the fractional changes \dmm\ and \daa\ are equal to
${\Delta v}/(c \Delta Q)$, 
where
$\Delta v = v_1 - v_2$ is the difference of the measured radial
velocities of these lines, and $\Delta Q = Q_2 - Q_1$ is the corresponding
difference between their sensitivity coefficients~\cite{L02,KL}. 

The $Q$ values of atomic transitions observed in quasar spectra 
are very small, $|Q| \ll 1$~\cite{Dz}.
Similar low sensitivity coefficients were calculated
for electro-vibro-rotational transitions in H$_2$ and CO (for references, see~\cite{KL}).
Small values of $Q$ and $\Delta Q$ put tough constraints on optical methods to
probe \daa\ and \dmm.
For instance, 
at  $\Delta \alpha/\alpha \sim 10^{-5}$, the required line position accuracy must
be $\sigma_v \la 0.25$ \kms\ in accord with the inequality~\cite{KL}:
$\sigma_v /c < (\Delta Q/\sqrt{2}) (\Delta \alpha/\alpha)$.
A typical error of the line center measurements of an
unsaturated absorption line in quasar spectra is about 1/10$th$ of the pixel size~\cite{L05}.
For high redshift objects,
the UV-Visual Echelle Spectrograph (UVES) at the ESO
Very Large Telescope (VLT) provides a pixel size $\Delta \lambda_{\rm pix} \sim 0.06$ \AA\ 
at $\lambda \sim 5000$ \AA, that is 
$\sigma_v \sim 0.5$ \kms, which is comparable to the velocity
offset due to a fractional change in $\alpha$ at the level of $10^{-5}$.
This shows that special care and
additional calibrations are required to probe \daa\ and \dmm\
at a level of $10^{-6}$ by optical methods.
Such measurements have been carried out at the VLT/UVES as described in the next section.

\section{VLT/UVES Large Program for testing fundamental physics}\label{sect-3}

The ESO Large Programme 185.A-0745 (2010-2013) was especially aimed at testing 
the hypothetical variability of physical constants~\cite{Mo,Ra,Bo,Ev}. 
Its prime goal was to study systematic errors in wavelength scales of quasar spectra. 
For this purpose quasars were observed almost simultaneously with 
bright asteroids, whose reflected
sunlight spectra contain many narrow features with positions
as accurate as a few \ms~\cite{Mol}.
Additionally, bright stars were observed through an iodine gas absorption
cell, providing a precise transfer function for part of the
wavelength range. 

As a result, there were revealed distortions of the wavelength scale with a
jig-saw pattern and peak-to-peak amplitude of several hundreds \ms\
along the echelle orders. 
The presence of long range wavelength dependent velocity drifts 
ranging between $\sim$0.5 and 1.0 \kms\ and showing
opposite sign as compared with the Keck/HIRES spectra of quasars
were found as well~\cite{WM}.

A stringent bound for \daa\ was obtained for the absorber at 
\zabs\ = 1.69 towards the quasar HE2217-2818~\cite{Mo}.
The fractional change of $\alpha$ in this system is 
\daa\ = $(1.3\pm2.4_{\rm stat} \pm 1.0_{\rm sys})\times10^{-6}$ if Al{\sc ii} $\lambda1670$ \AA\ 
and three Fe{\sc ii} transitions are used,
and \daa\ = $(1.1\pm2.6_{\rm stat})\times10^{-6}$ in a slightly different analysis with only 
Fe{\sc ii} transitions used. 
Together with another system observed with the UVES/VLT at \zabs\ = 1.58 towards HE0001-2340  
where \daa\ = $(-1.5\pm2.6_{\rm stat})\times10^{-6}$~\cite{Ag},
and eight HIRES/Keck quasar absorbers with the mean \daa\ = $(-0.1\pm2.6)\times10^{-6}$~\cite{Son},
these values are the tightest bounds to date on $\alpha$-variation at high redshifts.
As seen, they do not show any evidence for changes in $\alpha$ at the precision level 
of $\sim3\times10^{-6}$ (1$\sigma$ confidence level, C.L.).

For the electron-to-proton mass ratio the analysis of the H$_2$ absorption lines of the
\zabs\ = 2.40 damped Ly-$\alpha$ system towards HE0027-1836 yields
\dmm\ =$(2.5 \pm 8.1_{\rm stat} \pm 6.2_{\rm sys})\times10^{-6}$~\cite{Ra}.
When corrections to the wavelength dependent velocity drift are applied then
\dmm\ = $(7.6 \pm 8.1_{\rm stat} \pm 6.3_{\rm sys})\times10^{-6}$.
At higher redshift \zabs\ = 4.22 the analysis of H$_2$ absorption lines
in the spectrum of J1443+2724 gives 
\dmm\ = $(9.5 \pm 5.4_{\rm stat} \pm 5.3_{\rm sys})\times10^{-6}$~\cite{Bag}.
These results are consistent with a null $\mu$-variation at the $\sim2\times10^{-5}$ 
(1$\sigma$ C.L.) precision level
over a lookback time of $\approx 12.4$~Gyr (10\% of the age of the Universe today).

\section{Microwave and submillimeter molecular transitions}\label{sect-4}

Radio astronomical observations allow us to probe variation of the
fundamental constants on the cosmological time scale
at a level deeper than $10^{-5}$.
In the microwave range there are a good deal of molecular
transitions arising in Galactic and extragalactic sources.
Electronic, vibrational, and rotational energies in molecular
spectra are scaled as $E_{\rm el} : E_{\rm vib} : E_{\rm rot} = 1 : \mu^{1/2} : \mu$.
This means that the sensitivity coefficients for pure vibrational
and rotational transitions are equal to $Q_{\rm vib} = 0.5$
and $Q_{\rm rot} = 1$. 
Molecules have also fine
and hyperfine structures, $\Lambda$-doubling, hindered rotation,
accidental degeneracy between narrow close-lying levels
of different types and all of them have a specific dependence on
the physical constants. 
Some of these molecular transitions are $\sim100$ times more sensitive to variations
of $\mu$ and $\alpha$ than atomic and electro-vibro-rotational transitions of H$_2$ 
and CO which are detected in six quasar absorbers between $z = 1.6$ and 2.7~\cite{Not}.
In addition, positions of narrow molecular lines arising from cold dark clouds in the Milky Way disk
can be measured with uncertainties of $\sigma_v \la 0.01$ \kms\ \cite{L13},
that is, the resulting sensitivity in radio bands
is about three orders of magnitude higher as compared with optical spectra.

The molecular transitions with enhanced sensitivity coefficients which are 
the prime targets for testing the constancy of the fundamental constants by
radio astronomical methods were recently reviewed in~\cite{KL}. 
For instance, inversion transitions of ammonia NH$_3$~--- one of the 
most abundant molecules in the interstellar medium~--- 
have sensitivity coefficients $Q_\mu = 4.5$~\cite{Flam}.
This enhancement occurs due to the tunneling effect 
depending on the action $S$ which is proportional to $\mu^{-1}$:
the ground state tunneling frequency $\omega \propto e^{-S}$.
Observations of the NH$_3$(1,1) inversion line and five HC$_3$N rotational lines
at \zabs\ = 0.89 towards PKS1830-211~\cite{Hen}, 
as well as 
the inversion (NH$_3$) and rotational (CS, H$_2$CO) lines at \zabs\ = 0.69
towards B0218+357~\cite{Kan}
led to constraints ($1\sigma$ C.L.):
$|\Delta\mu/\mu| < 5\times10^{-7}$ and
$|\Delta\mu/\mu| < 1\times10^{-7}$, respectively.  

The second molecule which is 
extremely sensitive to $\mu$-variation and which is
observed in galactic and extragalactic molecular clouds is methanol CH$_3$OH. 
The sensitivity coefficients $Q_\mu$ for different transitions in CH$_3$OH
range from $-53$ to 42~\cite{Jan,L11}.
A distinctive feature of methanol is strong interaction
between the internal (hindered) and overall rotations.
Transitions, in which both the internal and overall rotation states are changed,
have strongly enhanced $Q_\mu$-factors.
However, the magnetic hyperfine structure of methanol transitions which was partly resolved in
laboratory measurements~\cite{Co} put natural restriction on the methanol method 
at the level of $\sim10^{-8}$ in \dmm\ tests.
The hyperfine coupling in methanol is due
to the well known magnetic spin-rotation and spin-spin
couplings leading to small line splittings of $\sim10$~kHz.
The large amplitude internal rotation may also lead to a less
known magnetic coupling~--- the so-called spin-torsion coupling~---
which has not yet been conclusively evidenced.

So far, methanol absorption lines were 
detected at \zabs\ = 0.89 in the gravitationally lensed system PKS1830-211~\cite{Mu}.
This system provides the most stringent limit on changes in $\mu$ over a lookback time
of $\approx 7.5$~Gyr: $|\Delta\mu/\mu| < 2\times10^{-7}$ ($1\sigma$ C.L.)~\cite{Kan15}.

Cold ($T_{\rm kin} \sim 10$~K) and dense 
($n_{{\scriptscriptstyle \rm H}_2} \sim 10^4$ cm$^{-3}$) molecular cores 
in the Milky Way disk are another perspective targets 
to probe $\mu$. 
The molecular cores are the ammonia emitters exhibiting some 
of the narrowest ($\Delta v \la 0.2$ \kms\ (FWHM)) lines ever observed~\cite{Ji,L16}.
The NH$_3$ linewidths $\Delta v$ of some of them correspond to a pure thermal broadening at 
a minimum gas temperature of $T_{\rm kin} \approx 8$~K coming
mainly from the heating by cosmic rays~\cite{Gol}.
A lifetime of molecular cores is $\sim10^{6-7}$~yr~\cite{Lee}, 
and they are located at regions with
different gravitational potentials.

A sample of molecular cores were studied with the Medicina 32-m, Nobeyama 45-m, and
Effelsberg 100-m telescopes in \cite{L13,L10a,L10b}.
The main result of these measurements is
the most stringent limit on $\mu$-variation 
for the period of $\sim10^{6-7}$~yr 
obtained by astronomical methods~\cite{L13}:
$|\Delta \mu/\mu| < 7\times10^{-9}$ ($1\sigma$ C.L.).
This upper limit is comparable with
the current constraint stemming from laboratory experiments, 
$\dot{\mu}/\mu < 6\times10^{-16}$~yr$^{-1}$~\cite{Fer}.

An independent test that $\alpha$ and $\mu$ may differ between
the high- and low-density environments of the Earth and the interstellar medium
was performed with CH and OH in~\cite{Tr13}.
In the Milky Way, the strongest limit to date 
on $\alpha$-variation  is $|\Delta\alpha/\alpha| < 1.4\times10^{-7}$ ($1\sigma$ C.L.).

Thus, the Einstein heuristic principle of LPI 
is validated all over the universe, that is, 
neither $\alpha$ at the level of $\sim {\it few}\times10^{-6}$, 
no $\mu$ at the level of $\sim {\it few}\times10^{-7}$  
deviates from its terrestrial value for the passed $10^{10}$ yr. 
Locally, no statistically
significant deviations of \dmm\ from zero were found at even more
deeper level of $\sim {\it few}\times10^{-9}$.
For the fine structure constant, such limit is $\sim10^{-7}$.

\section{Future prospects}\label{sect-5}

In previous sections we demonstrated that
the radio observations of NH$_3$ and CH$_3$OH lines are an order of
magnitude more sensitive to fractional changes in $\mu$ than the optical 
constraints derived from H$_2$.
However, at cosmological distances
there are only five radio molecular absorbers known 
and all of them are located at $z < 1$,
whereas H$_2$ lines are detected at redshifts $2 \la z \la 4$. 

As was emphasized in~\cite{KL},
the improvements in measurements of \daa\ and \dmm\
at the level of, respectively, $10^{-8}$ and $10^{-9}$,
can be achieved if two main requirements will be fulfilled: 
$(i)$ increasing precision of the laboratory measurements of
the rest frame frequencies of the most sensitive molecular
transitions, and 
$(ii)$ increasing sensitivity and spectral resolution of astronomical
observations.

The second requirement is expected to be realized in a couple of years when 
the Next Generation Very Large Array (ngVLA) will start regular operations~\cite{Bow}.
The ngVLA will provide ten times the effective collecting area of the
JVLA and ALMA, operating from 1~GHz to 115~GHz, with ten times longer
baselines (300~km).
The increased sensitivity of the ngVLT by an order
of magnitude over the VLA would allow discovery of new molecular absorbers at $z > 1$
and, thus, would extend the sample of targets suitable to  
test the EEP at early cosmological epochs. 

In optical sector,
the forthcoming generation of new optical telescopes such as
the Thirty Meter Telescope (TMT) and the European Extremely Large Telescope (E-ELT)
equipped by high-resolution ultra-stable spectrographs will significantly improve
the constancy limits of fundamental couplings. 
The future high precision optical measurements  
should achieve sensitivities of $\sim 10^{-7}$ for individual absorbers.  
Thanks to a large sample of absorption-line systems, 
a few times deeper limit is expected for the ensemble average. 

In spite of a far higher sensitivity of radio methods as compared to
that of next-generation optical facilities, the unresolved
(or partly resolved) magnetic hyperfine structure of molecular transitions 
prevents the radio measurements to achieve the accuracy better than $\sim10^{-9}$.

For example, the hyperfine structure of several transitions in methanol CH$_3$OH 
was recently recorded in the microwave domain 
using the Fourier transform microwave (FT-MW) spectrometer in Hannover
and the molecular beam FT-MW spectrometer in Lille~\cite{Co}.
With the line splitting of $\sim10$~kHz 
revealed in these laboratory studies, and the difference between the
sensitivity coefficients $\Delta Q_\mu \sim 10$ for the 48.372, 48.377, and 60.531~GHz 
methanol lines observed at
\zabs\ = 0.89 towards PKS1830-211~\cite{Kan15}, one finds the uncertainty of \dmm\
of about $3\times10^{-8}$, which is entirely caused by the unresolved hyperfine structure
of methanol lines.

It should be obvious that further progress in radio sector is in need of
accurate laboratory measurements of the rest frame molecular frequencies.
The required uncertainty of laboratory frequencies is $\la 1$~\ms.
There is currently a shortage of such data. 
Among molecules with high sensitivity coefficients to changes in $\mu$ and $\alpha$
only NH$_3$~\cite{Kuk67} and CH~\cite{Tr13,Tr14} transitions fulfill this requirement.

\section{Conclusions}\label{sect-6}

In this short review we highlighted the most important 
observational results which mark the frontier of most 
precise spectroscopic measurements of line positions in optical and radio sectors
aimed at different tests of the variation of fundamental physical constants
by astrophysical methods. 

Current null results from the VLT and Keck optical telescopes
as well as from different radio telescopes
validate the Einstein equivalence principle 
at a rather deep level of $\sim10^{-7}-10^{-6}$ for extragalactic sources,
and at $\sim10^{-8}$ within the Milky Way disk.
This is a tremendous step forward in experimental justification of basic principles of 
the general relativity and the standard model of particle physics as
compared with the first astrophysical constraint on $|\Delta\alpha/\alpha| < 3\times10^{-3}$
towards radio galaxy Cygnus~A ($z = 0.057$) obtained 60 years ago by Savedoff~\cite{Sa}.

It should to be emphasized that both optical and
radio methods complement each other and in future 
will provide independent tests of \daa\ and \dmm\ variability using 
the next-generation radio and optical telescopes.

\end{document}